\documentclass[aip, amsmath, amssymb, reprint]{revtex4-1}

\usepackage{graphicx}
\usepackage{dcolumn}
\usepackage{bm}

\usepackage[utf8]{inputenc}
\usepackage[T1]{fontenc}
\usepackage{mathptmx}
\usepackage{gensymb}

\begin{document}

\preprint{AIP/123-QED}

\title[]{Stability of bakeable capacitance diaphragm gauges}

\author{Julia Scherschligt}
\email{julia.scherschligt@nist.gov}
\author{Daniel Barker}
\author{Stephen Eckel}
\author{James Fedchak}
\author{Emmanuel Newsome}
 
\affiliation{ 
Sensor Sciences Division, National Institute of Standards and Technology, Gaithersburg, MD 20899
}

\date{\today}

\begin{abstract}
Here, we study the stability of bakeable capacitance diaphragm gauges.  We focus on their stability before and after a controlled series of bakes. We find that baking results in appreciable shifts of the zero offset, but note that these can easily be corrected at time of use. Linearity however cannot be corrected at time of use, so it is essential to understand its effect if one wishes to use similar gauges in a system that requires baking. For the gauges in this study, we find that baking introduces a minimal additional uncertainty, and that the total uncertainty can be kept to below about 0.3~\% at the $k=2$ confidence level ($95\%$). 

\end{abstract}


\maketitle

\section{Introduction}
Capacitance diaphragm gauges (CDGs) are the workhorse of pressure metrology around an atmosphere. They have been used extensively by the National Institute of Standards and Technology (NIST) and other national metrology institutes as transfer standards, notably for intercomparisons~\cite{Ricker2017,Miiller1999}. The characteristics that make them suitable for this purpose are their sensitivity and repeatability~\cite{Kojima2015Long-termGauges}, as well as their resistance to mechanical shock. Furthermore, because they are true sensors of pressure (as opposed to sensors of a proxy such as gas conductivity) they are gas species independent~\cite{Jousten2011}. 

Some commercial CDG  models consist of an all-metal construction (or nearly all-metal) and can therefore be can be baked at high temperatures.
Many applications of vacuum, particularly ultra- and extreme-high vacuum (UHV and XHV, respectively) require baking as part of system preparation.
Low temperature bakes in the range of 100~\degree C to 200~\degree C are routinely carried out on fully assembled vacuum systems in cases where volatile compounds need to be removed quickly, or where water or other substances with a high sticking coefficient need to be degassed.
Applications in the UHV or XHV typically require baking at much higher temperatures, to remove hydrogen dissolved in the stainless steel that comprises the bulk of most vacuum apparatuses.~\cite{Fedchak2021}  

Here, we study three bakeable CDGs and test the stability of their calibration before and after a bake.
The gauges under study are series 616A Baratron gauges produced by MKS instruments.             
These gauges are specified to be bakeable up to at least 400~\degree C, and operable up to at least 300~\degree C.
We test three gauges with nominal upper scale limits of 133 Pa, 1333 Pa, and 133322 Pa.
Herein we refer to these gauges as 1~torr, 10~torr, and  1000~torr CDGs, respectively.
Except were noted, all uncertainties are reported at the $k=1$ ($67\%$) confidence interval.
\footnote{Any mention of commercial products is for information only; it does not imply recommendation or endorsement by NIST nor does it imply that the products mentioned are necessarily the best available for the purpose.}

\section{Apparatus}
Before use, the CDGs were fired inside a custom-built vacuum furnace~\cite{Fedchak2018} for 33 days at approximately 425 \degree C to reduce hydrogen outgassing, following procedures developed in  references~\onlinecite{Sefa2017,Mamun2013}. The gauges were not tested prior to this high-temperature bake.  

\begin{figure}
    \centering
    \includegraphics{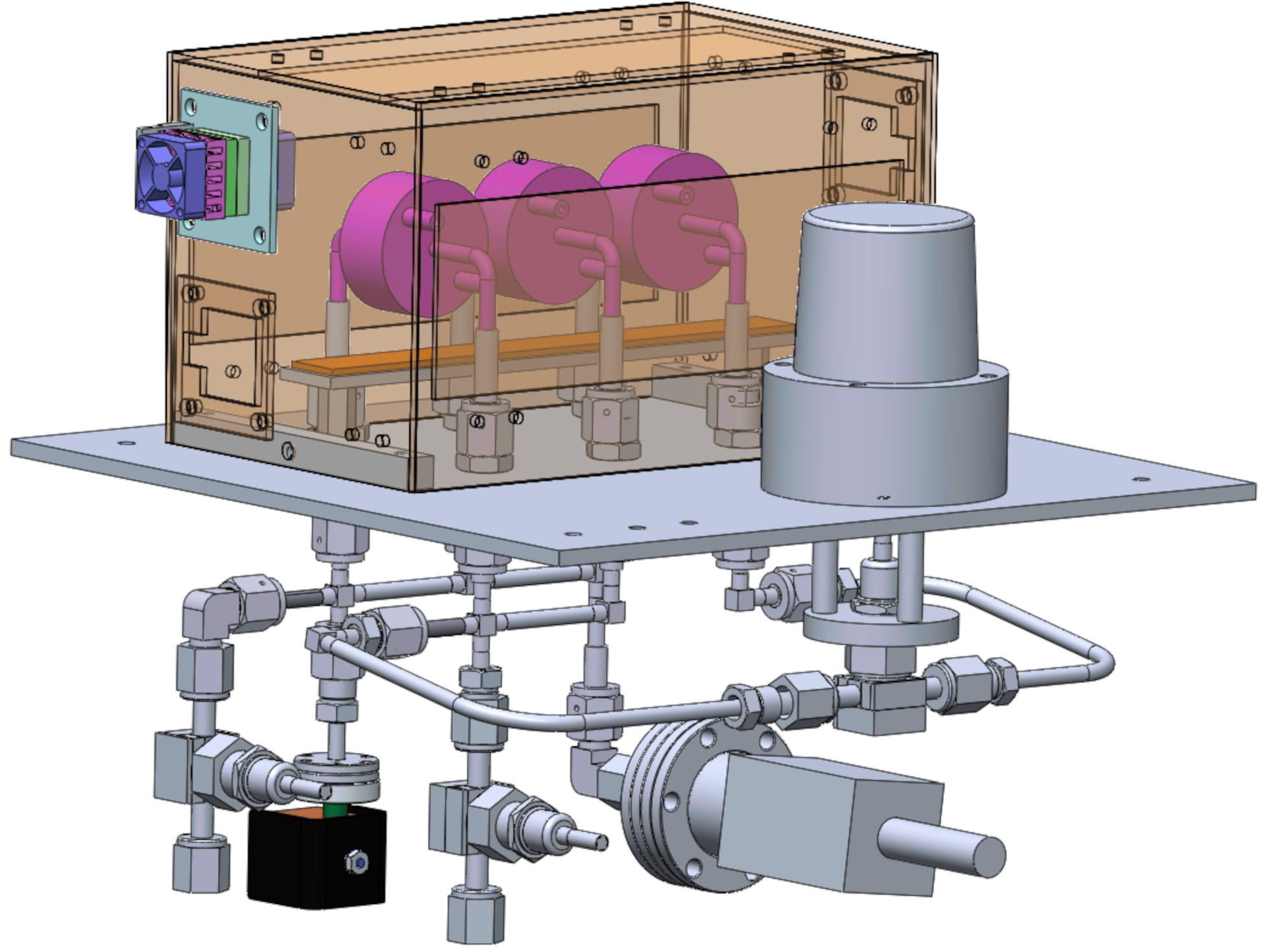}
    \caption{Bakeable CDGs (fuschia) housed inside a temperature controlled box (semi-transparent beige). The TEC unit for temperature control near room temperature protrudes from the left side of the box.}
    \label{fig:cozy_render}
\end{figure}

The CDGs are housed inside an aluminum box, 15~cm by 15~cm in by 25~cm in size, shown in Figure ~\ref{fig:cozy_render}.
The aluminum box is insulated on the inside by 2.5 cm thick Marinite board.  A custom-built TEC cooler unit is placed through one of the access ports and stabilizes the temperature inside the box to $\approx 25$~\degree C.
The gas handling manifold was designed to minimize volume, and is equipped with a pneumatic shut-off valve to isolate the 1~torr gauge from high pressure fills. The system is roughed out on both sides through a single access port.
Once sufficiently low pressures are achieved ($<10^{-3}$~Pa), the bypass valve to the reference side is closed and a 2.0~L/s ion pump is turned on.
The base pressure achieved on the reference side is typically of the order of $10^{-6}$~Pa.

\section{Comparison \& Results}
After vacuum firing and assembling, we performed an initial comparison of the three bakeable CDGs to a well-characterized set of three transfer standard CDGs, comprised of a 1~torr, a 10~torr and a 100~torr gauges and described in Ref.~\onlinecite{Ricker2017}.
The transfer standard CDGs are housed in a seperate temperature-controlled enclosure, and are calibrated using the NIST primary laboratory standard for pressure, the Ultrasonic Interferometer Manometer~\cite{Ricker2017}.
The initial comparison indicated a large difference in linear sensitivity between the two sets of gauges.

In particular, the bakeable 10~torr gauge read approximately $35$~\% larger than its transfer standard counterpart, and the 1000~torr gauge read approximately $54$~\% larger than its counterpart.
The difference in sensitivity was much larger than the manufacturer's specification.
Assuming the sensitivities were properly set at the factory, these initial discrepancies suggest that the high temperature 450~\degree C bake significantly changed the elastic properties of the Inconel membrane of the CDGs.
We adjusted the offset and span controls to bring them closer to alignment.
The discrepancy for the 1~torr gauges was small ($<5$~\%), and no adjustment was made to the factory settings.

\begin{figure}
    \centering
    \includegraphics{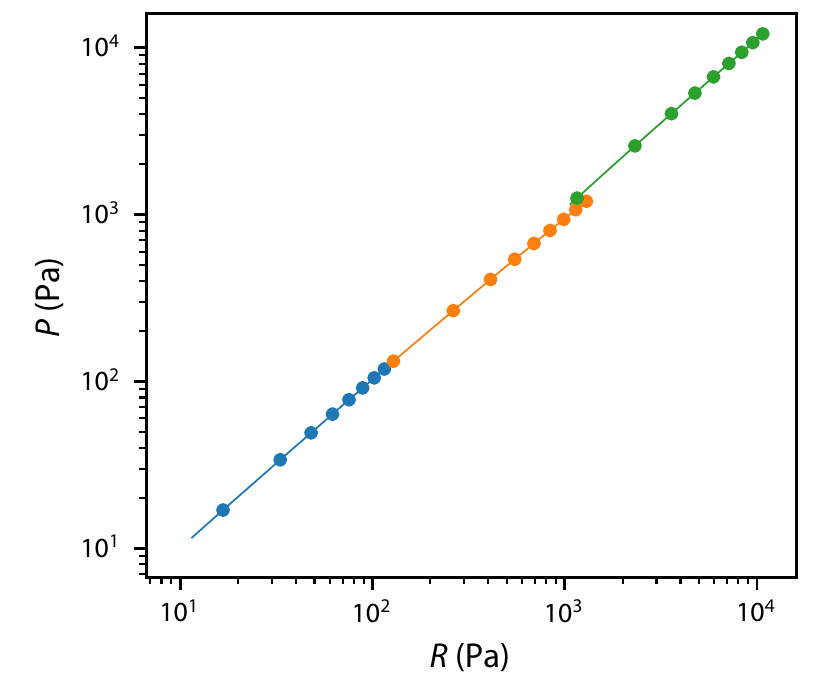}
    \caption{
    Pressure according to transfer standard $P$ vs. reading $R$, for a randomly selected run (\#42).
    The blue points correspond to data to calibrate the 1 torr gauge (gain X1), the orange to the 10 torr gauge (gain X1), and the green to the 1000 torr gauge (gain X0.1).
    Individual fits to \eqref{eq:single_run_fit} are shown as solid curves.
    }
    \label{fig:example_run}
\end{figure}

We then performed a subsequent comparison of the bakeable CDGs to our transfer standard following the procedure outlined below.
First, we evacuated both the transfer standard and the bakeable CDGs to $<0.4$~Pa and recorded pressures.
We tested the gauges at pressures covering the entire range of each gauge, selecting nine nominally equally spaced pressures to cover for each of three pressure decades (eg., 0.1, 0.2, 0.3, ... 0.9, 1, 2, etc.) for pressures up to 12~kPa. 
Pressure readings were recorded for all possible combinations of gauge, gain, and range that could access the set pressure.
For example, the 10~torr gauge at gain X0.1 has a full-scale pressure range of 1 torr (133 Pa), so its readings were recorded on the decade spanning 0.1 torr (13 Pa) to 0.9 torr (120 Pa). 
A full sweep through the pressure ranges is labeled a ``run''.
In most cases a run requires about 6 hours to complete. A typical run is shown in Figure~\ref{fig:example_run}.

Generally, we acquired 10 runs before discontinuing and baking the CDGs.
At this point, the TEC is removed and replaced with insulation, and internal heaters controlled by a thermocouple raise the temperature to 110~$^\circ$C at a rate of 1~$^\circ$C/min\footnote{During assembly, leaks were detected near the feed-through pins. These leaks were sealed with a solvent-free epoxy with a low vapor pressure resin rated for ultra-high vacuum, which limits our maximum bake out temperature to 110~\degree C.}.
After reaching 110~\degree C, the elevated temperature is maintained for anywhere between 3 to 5 days, at which point the system is again cooled to room temperature.
The TEC is reinserted, and the temperature is allowed to stabilize at 25~$^\circ$C for 24 hours before a new set of runs commences. 

To begin our analysis, for a given gauge each run is initially fit to a polynomial of the form
\begin{equation}
    \label{eq:single_run_fit}
    P = \sum_p c_p R^p,
\end{equation}
where $R$ is the pressure reading of the bakeable CDGs and $P$ is the true pressure as determined by the transfer standard.
The order of the polynomial used in the fit is three ($p=4$). 
Any points are rejected as outliers if their residual was more than eight standard deviations away from zero (the mean of the residuals).
The data are weighted by the estimated uncertainty in the transfer standard for the given pressure.

\begin{figure}
    \centering
    \includegraphics{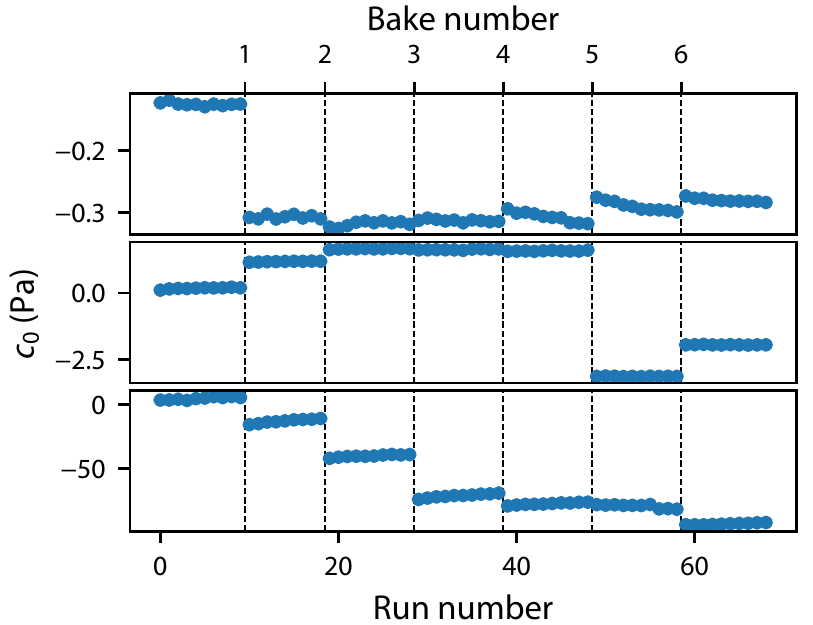}
    \caption{The offset $c_0$ fit coefficents versus run number for the 1 torr gauge, gain X1 (top); 10 torr gauge, gain X1 (middle); and 1000 torr gauge, gain X0.1 (bottom).  Bakes are shown as vertical dashed lines.  Error bars are present at the 1-$\sigma$ statistical uncertainty in the fit parameter, but are smaller than the data points.}
    \label{fig:c0_coefficients}
\end{figure}

\begin{figure}
    \centering
    \includegraphics{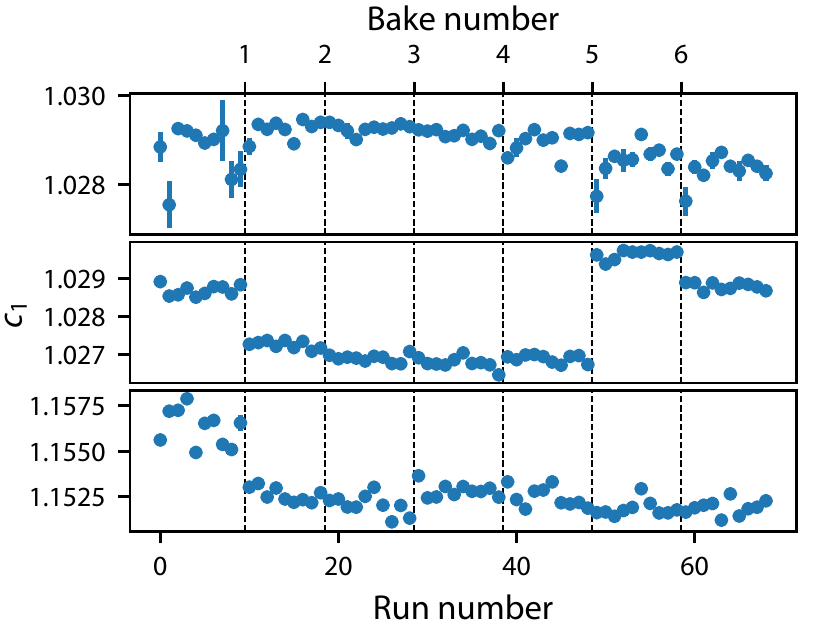}
    \caption{The linear sensitivity $c_1$ fit coefficents versus run number for the 1 torr gauge, gain X1 (top); 10 torr gauge, gain X1 (middle); and 1000 torr gauge, gain X0.1 (bottom).  Bakes are shown as vertical dashed lines.  Uncertainty bars are reported as $k=1$ uncertainty in the fit parameter.}
    \label{fig:c1_coefficients}
\end{figure}

Figures~\ref{fig:c0_coefficients} and~\ref{fig:c1_coefficients} show the resulting best fit coefficients $c_0$ and $c_1$ versus run number, respectively. 
For now, we ignore higher order coefficients.
Bakes are represented by vertical dashed lines.
Baking causes large discontinuities in the zero offset, the $c_0$ term, between several runs for the 10~torr and 1000~torr gauges.
For the 1~torr gauge, this discontinuity is only apparent after the first bake; all subsequent bakes maintain a relatively stable zero offset.
The discontinuities are less apparent in the linear sensitivity coefficient $c_1$.
After the first bake, the  $c_1$ coefficient appears to stabilize for 1~torr and 1000~torr gauges, and the 10~torr gauge experienced only a slight discontinuity in $c_1$ after the second bake.
Proceeding the fifth bake (between runs 47 and 48),
three additional tests were performed.
First, a mass of approximately 3~kg was intentionally dropped from a height of about 0.5~m
onto the platform on which the entire assembly rested, inducing a shock.
Second, the bellows connecting the CDGs to the test apparatus was disconnected and reconnected while keeping the CDGs under vacuum, potentially subjecting them to torque.
Third, the plumbing leading to the CDGs was wrapped with heater tape, insulated, and baked at 150~\degree C while baking the CDGs themselves at the standard 110~\degree C.
These additional tests do not appear to have affected the 1000~torr gauge or 1~torr gauges, while dramatically impacting the 10~torr gauge.
The 1~torr gauge appears to be experiencing a downward drift over time.
\begin{figure}
    \centering
    \includegraphics{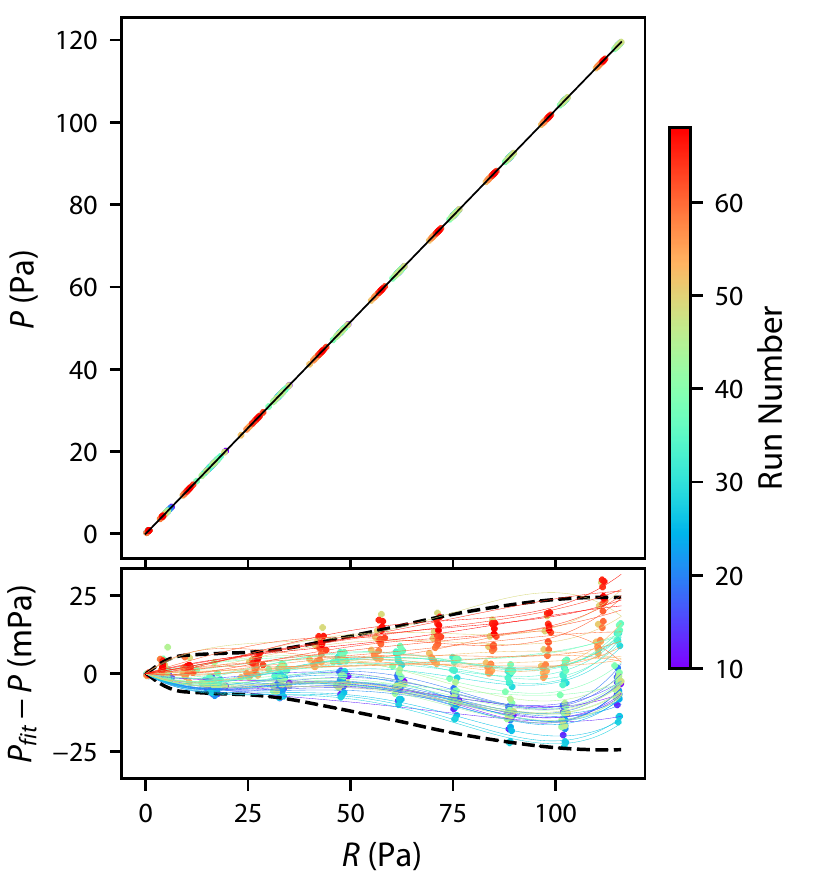}
    \caption{Example of the global polynomial fit for the 1~torr gauge at gain X1. (Top) Calibrated pressure vs. pressure reading $R$ for all runs after the 1st bake.  Dots show experimental data with color indicating run numbers and the solid black curve is the fit. Because actual pressure can differ somewhat from the setpoint, data from different run numbers are slightly dispersed.  (Bottom)  Residuals of the fit vs. pressure reading $R$. Colored curves indicate the difference of the single-run fit from the global one.  Dashed, black curves show the estimated $2$-$\sigma$ width of the residuals.}
    \label{fig:best_fit}
\end{figure}

While the zero offset is easily determined by evacuating the system to a pressure below the minimum sensitivity of the gauge, the repeatability of the linear sensitivity coefficient $c_1$ needs to be understood to use the gauge reliably~\cite{Boineau2018}.
Drifts in $c_1$ over time will result in uncertainty in the pressure reading.
(In fact, these drifts are the predominant source of uncertainty in our transfer standard.)
Excluding the runs before the first bake, the $c_1$ coefficients vary by about 0.08~\%, 0.2~\% and 0.1~\% for the 1~torr (X1 gain), 10~torr (X1~gain) and 1000~torr (X0.1~gain) gauges, respectively.
These variations are consistent with the long-term stability seen over 20 years of calibration of our transfer standard, which uses traditional unbakeable CDGs. For a description of the transfer standard, see Ref.~\onlinecite{Ricker2017}.
However, in order to fully quantify the long-term stability of the sensor, other terms must also be considered.
Because the best fits are polynomials, variations in the linear sensitivity coefficients are strongly correlated with the quadratic, cubic, and higher terms, and thus variations in $c_1$ do not by themselves fully characterize the fluctuations of the gauge.

To better quantify the variations between each calibration run, we repeat the fit procedure for each run with run number $r\geq 10$, i.e., excluding the data before the first bake.
Our global fit function uses the same polynomial coefficients as \eqref{eq:single_run_fit}, but with unique zero offsets for each run.
Specifically, the pressure read by a particular CDG during run $r$ is fit to
\begin{equation}
    \label{eq:multi_run_fit}
    P_r = c_{0,r} + \sum_{p\geq 1} c_p R_{r}^p
\end{equation}
where $R_r$ are the corresponding pressure readings and $c_{0,r}$ is the zero offset of run $r$.
All pressure readings $R_r$ and true pressures $P_r$ are fit together to obtain $c_p$ and $c_{0,r}$.
An example of this global fit for the 1~Torr gauge (X1 gain) is shown in Fig.~\ref{fig:best_fit}, along with the residuals.
The color of the points encodes the run number, and the colored curves in the residuals show the difference between the single-run fit and the global fit for a given run.
As seen in Fig.~\ref{fig:best_fit}, there is both a random component and drift over time (as indicated by the blue-to-red coloring from bottom to top).

\begin{figure}
    \centering
    \includegraphics{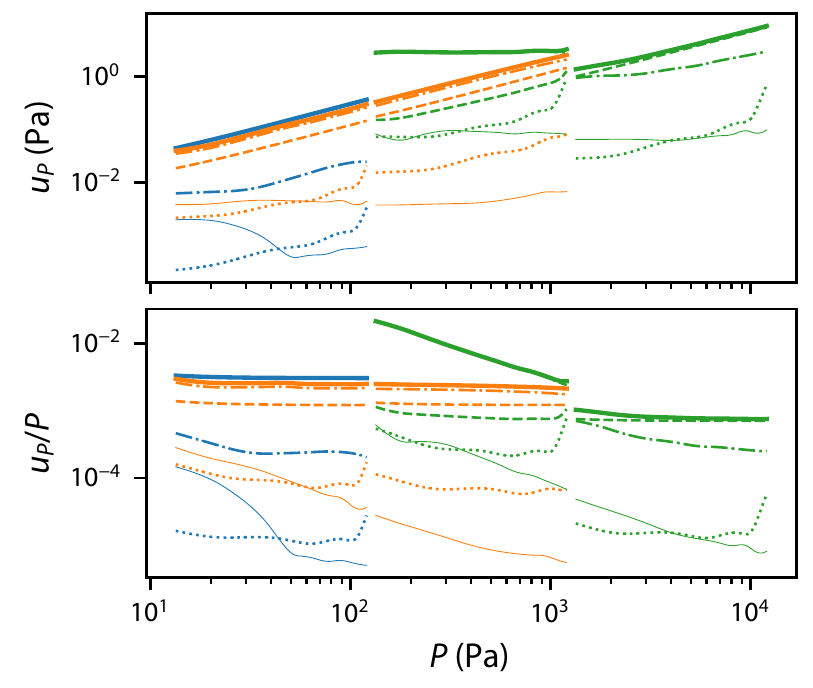}
    \caption{Uncertainties (top) and relative uncertainties (bottom) for the 1~torr (blue), 10~torr (orange), and 1000~torr (green) gauges vs. pressure $P$ between 13.3~Pa (0.1~torr) to 13.3~kPa (100 torr).  The thick solid lines denote the total uncertainty $u_P$, the dashed curves denote the uncertainty in the pressure transfer standard $u_\text{trans}$, the dashed-dot curves the uncertainty in the long-term stability $u_\text{lts}$, the dotted curves the uncertainty in the fit $u_\text{fit}$, and the thin solid curves the average uncertainty due to random fluctuations $u_{rdm}$ during a single calibration run. The uncertainties are $k=2$.}
    \label{fig:uncertainity}
\end{figure}

The black, dashed curves represent the statistical width of the residuals, given as twice the standard deviation ($k=2$ or $2$-$\sigma$).
The average ratio of the $2$-$\sigma$ width to the pressure is 0.04~\%, 0.2~\% and 0.04~\% for the 1~torr (X1 gain), 10~torr (X1~gain) and 1000~torr (X0.1~gain) gauges, respectively.
This ratio tends to be smaller than the variation in the individual fit $c_1$ parameters.
Again, this discrepancy between the standard deviation of the residuals and the scatter of the individually-fit $c_1$ coefficients is caused by neglecting correlations with the higher order polynomial coefficients.
The standard deviation of the residuals around the global fit provides an accurate measurement of the long-term stability of the gauge.
Technically, this scatter is the combination of the long-term drifts of the transfer standard and the long-term drifts of the bakeable CDGs.
However, for simplicity we assume that the transfer standard is perfectly stable during the duration of this experiment.
(By comparing the three CDGs in our transfer standard against each other, we can limit their calibration drift by $<0.04$~\% over the entire experiment.) 
Then, the scatter in the residuals allows us to estimate the uncertainty due to the long-term stability $u_\text{lts}$ in the operation of the CDGs, apart from the uncertainty in their calibration itself.
(We note that our assumption of a stable transfer standard will cause this uncertainty to be overestimated.)
In addition to the long term stability, there are several other sources of uncertainty in the operation of the CDGs. The full uncertainty in the pressure $u_P$ is given by
\begin{equation}
    \label{eq:uncertainity}
    u_P^2 = u_\text{cal}^2 + u_\text{lts}^2 + u_\text{rdm}^2 +  u_\text{zero}^2
\end{equation}
where $u_\text{cal}$ is the uncertainty in the gauges' calibration, $u_\text{lts}$ is the uncertainty in the long-term stability, quantified above, $u_\text{rdm}$ is the uncertainty due to random fluctuations during a measurement, and $u_\text{zero}$ is the uncertainty in the zero measurement.
Each of these uncertainty components is shown in Fig.~\ref{fig:uncertainity}.

The calibration's uncertainty $u_{cal}$ is itself comprised of three components, the uncertainty in the pressure of the transfer standard $u_\text{trans}$, the uncertainty in fit $u_\text{fit}$, and some random fluctuation uncertainity $u_\text{rdm}$.
The latter two components are by far the smallest, made small by the large amount of data that comprises each fit.
The uncertainty in the transfer standard is mostly dominated by the estimate of its long-term stability.

The remaining uncertainties, $u_\text{rdm}$ and $u_\text{zero}$ tend to be small by comparison.
These both need to be determined at time-of-use.
However, we include the $u_\text{rdm}$ component measured during the calibration in Fig.~\ref{fig:uncertainity} as an example.
For the calibration data, we recorded a point about every 5~s for 60~s, resulting in 12 independent measurements for a given gauge at a given pressure. 
With this scheme of taking data, $u_\text{rdm}$ is approximately 10 times smaller than $u_\text{lts}$.
The uncertainty in the zero will be a similar value, provided that the gauge is properly and routinely zeroed.
In summary, we expect that even with baking, the total uncertainty will be less than 0.3~\% for all the gauges and ranges tested.

\section{Conclusion}

Many applications in the vacuum require baking to remove water and other substances, so bakeable gauges are an essential component of such systems.
For applications that require the accuracy and precision of calibrated gauges, it is necessary to understand if baking changes the calibration, to what degree, and how to correct it.
This study of a particular set of bakeable gauges revealed that they can be made as stable as their non-bakeable counterparts. 
Indeed with all gauges of this type (CDGs) regardless of whether bakeable or not, temperature stability is key to maintaining the calibration.

Practical use of CDGs may subject them to other stresses that could affect their stability.
Future work may include studies of CDG stability under conditions of transport, plumbing modifications (for instance wrapping with heater tape, or repeated connecting and disconnecting), and accidental venting.
Furthermore, it might be beneficial to build a housing with better vibration isolation and to temperature stabilize the entire measurement system, rather than just the gauges (i.e. the parts shown in gray below the plate in Figure \ref{fig:cozy_render}).

\section*{Acknowledgments}
We thank Christopher Meyer, Eric Norrgard, and Weston Tew for technical assistance.  We thank Jacob Ricker for both technical assistance and useful discussions.

\bibliography{bakeablecdgs}

\end{document}